\documentclass[usenatbib]{mn2e_modmargin}
\usepackage{amsmath}
\usepackage{multirow}
\usepackage{url}
\usepackage{graphicx}
\usepackage{subfig}{
\usepackage{txfonts}
\usepackage{natbib}

\def\bnabla{\mbox{\boldmath $\nabla$}}
\def\half{{\textstyle{\frac12}}}
\def\cross{\times}
\def\machineepsilon{\varepsilon}%
\def\lisb{Lower is better.}%

\title{Symplectic integrators in the shearing sheet}
\author[H. Rein and S. Tremaine]{Hanno Rein\thanks{E-mail: rein@ias.edu} and Scott Tremaine\\
Institute for Advanced Study, 1 Einstein Drive, Princeton, NJ 08540, USA}
\begin{document}

\date{Submitted: 5 March 2011; Revised: 18 April 2011; Accepted: 19 April 2011 ; Published: - ;  }

\pagerange{\pageref{firstpage}--\pageref{lastpage}} \pubyear{2011}

\maketitle

\label{firstpage}

\begin{abstract}
     The shearing sheet is a model dynamical system that is
     used to study the small-scale dynamics of astrophysical
     disks. Numerical simulations of particle trajectories in the
     shearing sheet usually employ the leapfrog integrator, but this
     integrator performs poorly because of velocity-dependent
     (Coriolis) forces. We describe two new integrators for this
     purpose; both are symplectic, time-reversible and second-order
     accurate, and can easily be generalized to higher
     orders. Moreover, both integrators are exact when there are no
     small-scale forces such as mutual gravitational forces between
     disk particles. In numerical experiments these integrators have
     errors that are often several orders of magnitude smaller than
     competing methods. The first of our new integrators (SEI) is
     well-suited for disks in which the typical inter-particle
     separation is large compared to the particles' Hill radii (e.g.,
     planetary rings), and the second (SEKI) is designed for disks in
     which the particles are on bound orbits or the separation is smaller
     than the Hill radius (e.g., irregular satellites of the giant
     planets). 
\end{abstract}


\section{Introduction}
Hill's approximation, or the shearing sheet approximation, is an
essential tool for the study of the small-scale dynamics of
astrophysical disks \citep[for a general review of Hill's
approximation see, e.g.,][]{BinneyTremaine2008}. The method was originally devised by
\cite{Hill1878} to study the motion of the Moon, and has been applied
to galaxy disks
\citep[e.g.,][]{GoldreichLyndenBell1965,JulianToomre1966,GoldreichTremaine1978},
accretion disks \citep[e.g.,][]{HawleyBalbus1992,StoneGardiner2010}, planetary rings
\citep[e.g.,][]{WisdomTremaine1988,Salo1992,Richardson1994,Crida2010,ReinPapaloizou2010}
and planetesimal disks
\citep[e.g.,][]{Tanga2004,Johansen2009,BaiStone2010,ReinLesurLeinhardt2010}. Hill's
approximation is widely used in numerical simulations when it is
impossible to model an entire disk with adequate numerical resolution. 

This paper discusses numerical methods for following orbits in Hill's
approximation.  We first describe the relevant equations of motion in
Sect.~\ref{sec:hill}.  We then review existing integrators in
Sect.~\ref{sec:integrators}, and describe two new algorithms.  In
Sect.~\ref{sec:tests} we compare the convergence and performance of all
these integrators.  We summarize and discuss several generalizations in 
Sect.~\ref{sec:conclusions}.


\section{Hill's approximation}\label{sec:hill}

For simplicity we consider mainly the three-body problem, although
the results we describe are easy to generalize to other systems such
as planetary rings (see Sect. \ref{sec:conclusions}). Thus we follow
the motion of two nearby small bodies with masses $m_1$ and $m_2$ in
the gravitational field of a large body of mass $M\gg m_1,m_2$ (a more
careful version of this derivation is given by \citealt{HenonPetit1986}). The
two small bodies follow approximately the same orbit around the large
body, and we assume that this mean orbit is circular with semi-major axis $a$. The
angular speed of the mean orbit is then $\Omega=[G(M+m)/a^3]^{1/2}$
where $m=m_1+m_2$.  Formally, to derive Hill's equations of motion, one lets
$m/M$ shrink to zero while assuming that the mass ratio $m_1/m_2$ is fixed 
and the separation between the
two small bodies is of order $(m/M)^{1/3}$.

Let us adopt a local right-handed Cartesian coordinate system with its
origin at $m_1$, rotating uniformly with angular velocity $\Omega$.
The unit vector $\mathbf{e}_x$ points away from the larger mass $M$,
the unit vector $\mathbf{e}_y$ points in the direction of motion along
the mean orbit, and the unit vector $\mathbf{e}_z$ points in the
direction of the angular velocity vector of the mean orbit.  Hill's
equations of motion for the relative position
$\mathbf{r}=\mathbf{r}_2-\mathbf{r}_1$ are then
\begin{align}
\mathbf{\ddot r}= -2\Omega\; \mathbf{e}_z \cross \mathbf{\dot r} + 3 \Omega^2 (\mathbf{r \cdot e}_x) \,\mathbf{e}_x -\Omega^2 (\mathbf{r \cdot e}_z) \,\mathbf{e}_z+\mathbf{f},\label{eq:hill}
\end{align} 
where the force exerted by $m_1$ on $m_2$ is
$\mathbf{f}=-\bnabla\Phi$ with
$\Phi(\mathbf{r})=Gm/|\mathbf{r}|$. 
The Hamiltonian corresponding to these equations of motion is
\begin{align}
H(\mathbf{r},\mathbf{p})&=
\half\mathbf{p}^2 + \Omega \left(\mathbf{p}\cross\mathbf{r}\right)\cdot\mathbf{e}_z+\half \Omega^2\left[\mathbf{r}^2 -3\left(\mathbf{r}\cdot\mathbf{e}_x\right)^2\right]+\Phi(\mathbf{r})\nonumber\\
&\equiv H_0(\mathbf{r},\mathbf{p})+ \Phi(\mathbf{r}), \label{eq:hamiltonian}
\end{align}
where $\mathbf{p}=\mathbf{\dot r}-\Omega\,\mathbf{r}\cross\mathbf{e}_z$ 
is the canonical momentum conjugate to~$\mathbf{r}$.

Note that when $\mathbf{f}=\mathbf{0}$ the equations of motion (\ref{eq:hill})
are invariant under the shear transformation 
\begin{equation}
\mathbf{r}\rightarrow \mathbf{r}+c_x\left(\mathbf{e}_x-\tfrac32\Omega\mathbf{e}_yt\right)+c_y\mathbf{e}_y
\label{eq:shear}
\end{equation}
for arbitrary constants $c_x$ and $c_y$. This invariance allows the use of
periodic shearing boundary conditions for the study of the local
dynamics of disks and is one reason why Hill's approximation is so
useful.

Alternatively, we can work in a non-rotating but accelerated reference
frame with origin at $m_1$. If the relative position in this frame is
$\mathbf{R}=(X,Y,Z)$ and the $(x,y,z)$ and $(X,Y,Z)$ reference frames
coincide at $t=0$, then
\begin{align}
\mathbf{\ddot R}= \Omega^2\left( [\mathbf{r \cdot
    e}_x(t)]\,\mathbf{e}_x(t)   -\mathbf{R}\right)+\mathbf{f},\label{eq:hill_inertial} 
\end{align} 
with $\mathbf{e}_x(t)=\cos(\Omega t)\mathbf{e}_X+\sin(\Omega
t)\mathbf{e}_Y$. The Hamiltonian is
\begin{align}
H(\mathbf{R},\mathbf{P},t)=\half\mathbf{P}^2+\half\Omega^2\left\{\mathbf{R}^2-3[\mathbf{R}\cdot\mathbf{e}_x(t)]^2\right\}+\Phi(\mathbf{R})
\end{align}
where $\Phi(\mathbf{R})=\Phi(\mathbf{r})=-Gm/|\mathbf{r}|$ since $|\mathbf{R}|=|\mathbf{r}|$.

The physical interpretation of Eqs.~(\ref{eq:hill}) and
(\ref{eq:hamiltonian}), and the optimum choice of integrator, depend
on the relative size of the terms on the right side.  This can be
parametrized using the Hill radius,
$r_{\text{Hill}} \equiv(\frac13{Gm}/{\Omega^2})^{1/3}$,
which sets the relevant scale in the problem.

\begin{itemize}
\setlength{\itemindent}{0em}
\setlength{\leftskip}{1em}
\setlength{\parskip}{1em}
\item If the mass of the small bodies is negligible ($m\rightarrow0,
  r_{\text{Hill}}\rightarrow0$) their relative motion is simply epicyclic
  motion, which can be solved exactly \citep[see e.g.][or
  below]{HenonPetit1986,BinneyTremaine2008}. If $m$ is sufficiently
  small, the motion can be regarded as a perturbed
  epicycle. Quantitatively, this requires that one or more of the
  first three terms on the right side of Eq.~(\ref{eq:hill}) is much
  larger than $|\mathbf{f}|=Gm/r^2$, or
\begin{equation}
\frac{r}{r_{\text{Hill}}} \gg \mbox{min}\,\left[1,(\Omega
  r_{\text{Hill}}/v)^{1/2}\right],
\label{eq:approx}
\end{equation}
where $v$ is the velocity. 
One example is a planetary ring with semi-major axis $a$ much less than
the Roche limit $a_R=2.46[3M/(4\pi \rho_p)]^{1/3}$; here $M$ is the planet mass, $m_1$ and $m_2$ are the
masses of the ring particles, and $\rho_p$ is the ring-particle
density. For two particles the separation between them cannot be
less than twice their radius, which implies that
$r/r_{\text{Hill}}\gtrsim a_R/a\gg1$.

\item If the force due to the mutual gravity of the small masses is
large, we can interpret the solution to Eq.~(\ref{eq:hill}) as a
Keplerian orbit perturbed by Coriolis and tidal forces (the terms
proportional to $\Omega$ and $\Omega^2$ respectively). For a circular
orbit of semi-major axis $r$ the ratio of these perturbing forces to
the Kepler force $Gm/r^2$ is $(r/r_{\text{Hill}})^{3/2}$ and
  $(r/r_{\text{Hill}})^3$ respectively. One example
is the irregular satellites of the giant planets of the solar system,
which typically have $r/r_{\text{Hill}}\simeq 0.1$--0.5. 
\end{itemize}

\section{Integrators}\label{sec:integrators}

\subsection{Leapfrog integrator}\label{sec:leapf}

Many N-body simulations, in Hill's approximation and other contexts,
use the standard leapfrog integrator
\citep[e.g.,][]{Springel2005,BinneyTremaine2008}. However, as we shall
show below, leapfrog does not work well in Hill's approximation
because there are velocity-dependent forces.
Leapfrog is best suited for integrating equations of motion of the
form
\begin{align}
\mathbf{\ddot r}= - \bnabla\Phi(\mathbf{r})\label{eq:motionsimple}
\end{align} 
where $\Phi(\mathbf{r})$ is a (in general time-dependent) potential. Note that Eq.~(\ref{eq:hill}) 
can not be written in this form. The Hamiltonian corresponding to Eq.~(\ref{eq:motionsimple}) is 
\begin{align}
H(\mathbf{r},\mathbf{p}) =\half\mathbf{p}^2+\Phi(\mathbf{r})\equiv
  H_{\text{Kin}}(\mathbf{p})
  +\Phi(\mathbf{r}). \label{eq:hamiltonianpotential} 
\end{align}
We assume that we have the position and velocity\footnote{For the
  Hamiltonian in Eq.~(\ref{eq:hamiltonianpotential}) the canonical
  momentum $\mathbf{p}$ is the same as the time derivative
  of the position (the velocity). This is not the case for the Hamiltonian in Hill's
  approximation, Eq.~(\ref{eq:hamiltonian}).}  of a particle
at time $t^n$, $\mathbf{r}^n\equiv \mathbf{r}(t^n)$ and
$\mathbf{v}^n\equiv\mathbf{\dot r}(t^n)=\mathbf{p}(t^n)$.  The goal is
to approximate the new position and velocity at time
$t^{n+1}=t^n+\Delta t$.  Leapfrog is usually written as a chain of
three operators, labeled Kick, Drift, Kick, applied
successively\footnote{One can also apply the operators in a different
  order: Drift, Kick, Drift.  This does not change any conclusion in
  this paper.}:
\begin{align}
\begin{array}{llll}
\quad\mathbf{v}^{n+1/2} 	&= \mathbf{v}^{n} 	&+\, \frac12 \Delta t\;\;\mathbf{f}^n	&\quad\quad\quad\quad\quad\quad\quad\quad\quad\text{Kick}\\ 
\quad\mathbf{r}^{n+1} 	&= \mathbf{r}^n 	&+\, \Delta t\,\mathbf{v}^{n+1/2} 		&\quad\quad\quad\quad\quad\quad\quad\quad\quad\text{Drift}\\
\quad\mathbf{v}^{n+1} 	&= \mathbf{v}^{n+1/2} 	&+\, \frac12 \Delta t\;\;\mathbf{f}^{n+1}  	&\quad\quad\quad\quad\quad\quad\quad\quad\quad\text{Kick,}\\ 
\end{array} \nonumber \hspace*{5cm} 
\end{align}
where $\mathbf{f}^n=-\bnabla\Phi(\mathbf{r}^n)$.  The kick
operator corresponds to following the trajectory exactly under the
influence of the Hamiltonian $\Phi(\mathbf{r})$, the potential energy,
and the drift operator corresponds to the Hamiltonian
$H_{\text{Kin}}(\mathbf{p})$, the kinetic energy.  Thus if we denote
the operator for the exact evolution of a trajectory for a time-step $\Delta t$ under an
arbitrary Hamiltonian $H$ by $\hat H(\Delta t)$, a single leapfrog step is
the operator $\hat H_{\text{Kin}}(\tfrac12 \Delta t)\hat \Phi(\Delta t) \hat
H_{\text{Kin}}(\tfrac12 \Delta t)$. Each of these
operators is symplectic since they are governed by a Hamiltonian,
which implies that the leapfrog operator is also symplectic \citep{SahaTremaine1992}. Moreover
leapfrog is time-reversible and second-order, that is, the error after
a single time-step is $O(\Delta t^3)$. 

When leapfrog is used in Hill's approximation, to integrate the
equations of motion governed by Eq.~(\ref{eq:hill}) rather than Eq.~(\ref{eq:motionsimple}), the additional terms on the right
side are added in the kick step, since they change the velocity. Thus
we have (partly in component notation, for clarity): 
\begin{align}
\begin{array}{llll}
\quad v_x^{n+1/2} 	&=& v_x^{n} + \frac12 \Delta t \left(3\Omega^2\, x^n + 2\Omega\,v_y^n +f_x^n\right)  \\
\quad v_y^{n+1/2}	&=& v_y^{n} + \frac12 \Delta t \left(-2\Omega\,v_x^n +f_y^n\right) &\text{Kick}\\
\quad v_z^{n+1/2}	&=& v_z^{n} + \frac12 \Delta t \left(-\Omega^2\,z^n +f_z^n\right)\\\\
\quad \mathbf{r}^{n+1}&=& \mathbf{r}^n + \Delta t\; \mathbf{v}^{n+1/2} &\multirow{1}{*}{\text{Drift}}\\\\
\quad v_x^{n+1}	&=& v_x^{n+1/2} + \frac12 \Delta t \left(3\Omega^2\, x^{n+1} + 2\Omega\,v_y^{n+1/2} +f_x^{n+1}\right) \\
\quad v_y^{n+1}	&=& v_y^{n+1/2} + \frac12 \Delta t \left(-2\Omega\,v_x^{n+1/2} +f_y^{n+1}\right)&\text{Kick}\\
\quad v_z^{n+1}	&=& v_z^{n+1/2} + \frac12 \Delta t \left(-\Omega^2\,z^{n+1} +f_z^{n+1}\right).
\end{array} \nonumber \hspace*{5cm} 
\end{align}
However, because of the velocity-dependent Coriolis force on the right
side, in this form leapfrog loses most of its desirable properties:
(i) it is neither symplectic nor time-reversible; (ii) it is only
first-order accurate rather than second-order, that is, the error
after a single time-step is $O(\Delta t^2)$; (iii) it leads to secular
drifts in quantities that should be conserved. 


In principle these difficulties could be avoided by using leapfrog to
integrate the equations of motion in the form
(\ref{eq:hill_inertial}), which have no velocity-dependent
forces so leapfrog is symplectic, time-reversible, and
second-order. We do not pursue this option because (i) these equations
are not invariant under the shear transformation, Eq.~(\ref{eq:shear}),
which limits their usefulness for the study of disks and rings; (ii)
the method described in Sect.\ \ref{sec:eei} below is better. 

There are schemes for constructing symplectic integrators for
arbitrary Hamiltonians, but these are generally implicit and often
algebraically complicated, at least for high-order schemes. The first
explicit symplectic integrator for trajectories in Hill's
approximation is due to \cite{Heggie2001}. However, Heggie's
integrator is not time-reversible and is only first-order; we will not
discuss it further because our numerical experiments show that it is
not competitive with the integrators described below.

\subsection{Modified leapfrog integrator}\label{sec:modlf}
One can improve the standard leapfrog algorithm by using 
a predictor step to approximate the velocity-dependent forces at the end of the time-step.
This leads to the following scheme
\begin{align}
\begin{array}{llll}
\quad v_x^{n+1/2} 	&=& v_x^{n} + \frac12 \Delta t \left(3\Omega^2\, x^n + 2\Omega\,v_y^n +f_x^n\right)  \\
\quad v_y^{n+1/2}	&=& v_y^{n} + \frac12 \Delta t \left(-2\Omega\,v_x^n +f_y^n\right) &\text{Kick}\\
\quad v_z^{n+1/2}	&=& v_z^{n} + \frac12 \Delta t \left(-\Omega^2\,z^n +f_z^n\right)\\\\
\quad \bar v_x^{n+1} 	&=& v_x^{n} + \Delta t \left(3\Omega^2\, x^n + 2\Omega\,v_y^n +f_x^n\right)  \\
\quad \bar v_y^{n+1}	&=& v_y^{n} + \Delta t \left(-2\Omega\,v_x^n +f_y^n\right) &\multirow{1}{*}{\text{Drift}} \\
\quad \mathbf{r}^{n+1}&=& \mathbf{r}^n + \Delta t\; \mathbf{v}^{n+1/2}\\\\
\quad v_x^{n+1}	&=& v_x^{n+1/2} + \frac12 \Delta t \left(3\Omega^2\, x^{n+1} + 2\Omega\,\bar v_y^{n} +f_x^{n+1}\right) \\
\quad v_y^{n+1}	&=& v_y^{n+1/2} + \frac12 \Delta t \left(-2\Omega\,\bar v_x^{n} +f_y^{n+1}\right)&\text{Kick}\\
\quad v_z^{n+1}	&=& v_z^{n+1/2} + \frac12 \Delta t \left(-\Omega^2\,z^{n+1} +f_z^{n+1}\right).
\end{array} \nonumber \hspace*{5cm} 
\end{align}
Here $\bar v^{n+1}$ is the predicted value of the velocity at time
$t+\Delta t$.  This integrator is second-order, one order higher than
standard leapfrog, but is neither time-reversible nor symplectic.
This method is implemented in the particle codes Gasoline and PkdGRAV
\citep{Wadsley2004}.

\subsection{Symmetrized leapfrog}\label{sec:lfsym}
\cite{MikkolaMerritt2006} describe a simple algorithm that converts any
one-step integrator to a time-reversible integrator. We have applied
the Mikkola--Merritt algorithm to the standard (first-order) leapfrog integrator of
Sect.\ \ref{sec:leapf}, thereby upgrading it to a
time-reversible and second-order (but not symplectic)
integrator. Time-reversibility endows integrators with most of the
same desirable properties as symplecticity.

The tests described below show that symmetrized leapfrog is not
competitive with some of the other integrators discussed here. However,
this elegant integrator could be profitably applied to systems
described by more complicated time-reversible Hamiltonians and also
systems that are not governed by a Hamiltonian at all.

\subsection{Quinn et al.\ integrator}\label{sec:quinn}

Recently, \cite{Quinn2010} described an integrator that exhibits the
desirable features of leapfrog despite the presence of
velocity-dependent forces in Hill's approximation. In particular, the
Quinn et al.\ integrator is symplectic, time-reversible, and accurate
to second order.  We refer the reader to the original paper for a derivation.
Here, we merely list the final algorithm, which can also be written as
three operators, Kick, Drift, Kick\footnote{The Drift step is the same
  as in standard leapfrog, but the Kick step is modified.}:
\begin{align}
\begin{array}{llll}
\quad v_x^{n+1/4}	&=& v_x^n - \frac 12 \Delta t \left( \Omega^2 x^n - f_x^n\right) & \multirow{5}{*}{\quad\quad\quad\quad\text{Kick}}\\
\quad P_y^{n}		&=& v_y^n + 2\Omega x^n + \frac12 \,\Delta t\, f_y^n\\
\quad v_x^{n+1/2} 	&=& v_x^{n+1/4} +\Delta t\,\Omega P_y^n \\
\quad v_y^{n+1/2}	&=& P_y^n - \Omega x^n - \Omega\left(x^n+\Delta t\;v_x^{n+1/2}\right)\\
\quad v_z^{n+1/2}	&=& v_z^{n} + \frac12 \Delta t \left(-\Omega^2\,z^n +f_z^n\right)\\\\
\quad \mathbf{r}^{n+1}&=& \mathbf{r}^n + \Delta t\; \mathbf{v}^{n+1/2} &\quad\quad\quad\multirow{1}{*}{\quad\text{Drift}}\\\\
\quad v_x^{n+3/4} 	&=& v_x^{n+1/2} + \Delta t\;\Omega\,P_y^n& \multirow{4}{*}{\quad\quad\quad\quad\text{Kick}}\\
\quad v_x^{n+1}	&=& v_x^{n+3/4} -\frac12 \Delta t \,\left(\Omega^2\,x^{n+1}-f_x^{n+1}\right) \\
\quad v_y^{n+1}	&=& P_y^n - 2\Omega\,x^{n+1} + \frac12 \Delta t f_y^{n+1}\\
\quad v_z^{n+1}	&=& v_z^{n+1/2} + \frac12 \Delta t \left(-\Omega^2\,z^{n+1} +f_z^{n+1}\right).
\end{array} \nonumber\hspace*{5cm} 
\end{align}

\subsection{Symplectic epicycle integrator (SEI)} \label{sec:eei}

Mixed variable symplectic (MVS) schemes such as the Wisdom-Holman
integrator have become the method of choice for long-term integrations
of planetary orbits \citep{WisdomHolman1991,Kinoshita1991}.  Like
leapfrog, MVS schemes split the Hamiltonian into two parts,
$H(\mathbf{r},\mathbf{p}) =H_A(\mathbf{r},\mathbf{p})
+H_B(\mathbf{r},\mathbf{p})$, each of which is analytically
integrable. First, the trajectory is advanced under the influence of
$H_B$ for half a time-step, then $H_A$ for a full time-step, then
$H_B$ again. In contrast to leapfrog, where $H_A$ and $H_B$ are the
kinetic and potential energy (cf.\ Eq.~\ref{eq:hamiltonianpotential}), in MVS schemes $H_A$ and $H_B$ are
chosen so that $|H_B|\ll|H_A|$. Thus in the planetary case $H_A$ is
chosen to be the Kepler Hamiltonian while $H_B$ represents the small
gravitational forces from other planets. MVS integrators are
symplectic (since the trajectory is advanced by a sequence of
Hamiltonian maps) and time-reversible, and the error per time-step is
$O(\Delta t^3)O(H_B/H_A)$.

Interestingly, it is even easier to derive an MVS integrator for
Hill's equations of motion than for Keplerian motion.  As we show
below, it is possible to solve for the epicyclic motion, that is the
motion that is governed by the Hamiltonian
$H_0(\mathbf{r},\mathbf{p})$ in Eq.~(\ref{eq:hill}), in closed form
with almost no computational effort \citep[see
also][]{HenonPetit1986}.  We use this result to construct two new MVS
integrators, one each for the two cases given at the end of Sect.\
\ref{sec:hill}: systems in which the forces $\mathbf{f}$ due to the
gravity of the small bodies or other sources are relatively small (this
section), and systems in which the gravity between a single pair of bodies
dominates their motion (Sect. \ref{sec:wheei}).

We first solve the equations of motion governed by
$H_0(\mathbf{r},\mathbf{p})$.  To do that we shift the particle to a
coordinate system where the particle's center of epicyclic motion is
at the origin. We then do a rotation to account for the evolution in
the epicycle. Finally we shift back and account for the shear.

Using the same notation as above\footnote{Note that $v$ is the
  velocity, not the canonical momentum.}, the center of epicyclic
motion of a particle is at
\begin{align}
x_0^n		&= 2 v_y^n\,\Omega^{-1}+4 x^n \label{eq:epia}\\
y_0^n		&= y^n-2v_x^n\,\Omega^{-1}.  \nonumber\\
\intertext{We can then define}
x_s^n		&= \Omega\,(x^n - x_0^n) \\
y_s^n		&= \tfrac12\Omega(y^n - y_0^n). \nonumber\\
\intertext{The evolution of these quantities during one time-step $\Delta t$ can be written as a rotation around the origin with an angle $\Omega \Delta t$, }
x_s^{n+1}	&= x_s^n \cos(\Omega\, \Delta t)  		+ y_s^n \sin(\Omega\, \Delta t)\label{eq:SEI2}\\
y_s^{n+1}	&= -x_s^n \sin(\Omega\, \Delta t) 		+ y_s^n \cos(\Omega\, \Delta t). \nonumber
\intertext{Now we have only to undo the previous shift to the center of epicyclic motion and account for the shear to get the position and velocity at the new time $t^n+\Delta t$: }
x^{n+1} 	&= x_s^{n+1}\, \Omega^{-1}+x_0^n\, \label{eq:fff} \\
y^{n+1} 	&= 2 y_s^{n+1} \, \Omega^{-1}+y_0^n - \tfrac32 x_0^n\,\Omega \,\Delta t \nonumber\\\nonumber
v_x^{n+1} 	&= y_s^{n+1} \\ \nonumber
v_y^{n+1} 	&= -2 x_s^{n+1}  - \tfrac32 x_0^n \,\Omega.
\intertext{The integration of the vertical motion can also be described by a rotation, so that the new vertical position and velocity at time $t^n+\Delta t$ are given by}
z^{n+1} 	&= z^n \cos(\Omega\, \Delta t) 		+ v_z^n \Omega^{-1} \sin(\Omega\, \Delta t)\label{eq:SEI3}\\
v_z^{n+1} 	&= -z^n\Omega \sin(\Omega\, \Delta t) 	+ v_z^n  \cos(\Omega\, \Delta t).  \nonumber \quad\quad\quad\quad\quad\quad\quad
\end{align}
In some uses of Hill's approximation such as galactic disks, the
epicycle and vertical frequencies may differ from the azimuthal
frequency $\Omega$, but this generalization is easy to
incorporate. The operator corresponding to the steps
(\ref{eq:epia})--(\ref{eq:SEI3}) may be written in our notation as
$\hat H_0(\Delta t)$. 

Note that no function evaluation had to be performed during the entire
step (i.e., there is no call to \texttt{sqrt()}). The sines and
cosines appearing in the above equations are constant and the same for
all particles. They can be pre-calculated at the beginning of the
time-step or even at the beginning of the simulation if the time-step is
fixed. All other operations are additions and multiplications. No
significant additional storage is needed when there are many
particles. Also note that the integrator can be completely described
by translations and rotations, making it an attractive choice for
programs running on graphic processors (GPUs).

In long, high-accuracy integrations, round-off errors in the rotations
in steps (\ref{eq:SEI2}) and (\ref{eq:SEI3}) can cause a linear drift
in energy and other integrals of motion, at a rate $|\Delta E/E| \sim
\machineepsilon \cdot(t/\Delta t)$, where $\machineepsilon$ is the machine precision,
typically $2^{-53}$ for double-precision arithmetic. An elegant
solution to this problem is described in Appendix~\ref{app:rot}.

An MVS integrator that includes additional forces due to a potential
$\Phi(\mathbf{r})$ may then be written as
\begin{align}
\hat H_{\text{SEI}}(\Delta t) &= \hat H_{\text{0}}(\tfrac12\Delta t)\;
\hat \Phi(\Delta t)\; \hat H_{\text{0}}(\tfrac12\Delta t),
\label{eq:hamsei}
\end{align}
where as usual $\hat\Phi(\Delta t)$ represents the kick step
\begin{equation}
\mathbf{v}^{n+1}=\mathbf{v}^n-\Delta t\;\bnabla\Phi\left(\mathbf{r}^{n+1/2}\right).
\end{equation}
This integrator is symplectic, time-reversible, and second-order, and
in contrast to the other integrators we have discussed so far, becomes
exact as $\bnabla\Phi\rightarrow 0$. More precisely, if the gravitational
potential $\Phi$ is $O(\epsilon)$ then the error of the SEI integrator
after a single time-step is $O(\epsilon\;\Delta t^3)$, while the error of the
Quinn et al.\ integrator is $O(\Delta t^3)$.

As pointed out by \cite{Quinn2010}, numerical codes that
implement collision detection usually assume that particles move along
straight lines. In that case collision detection can be done exactly
(although it is often done approximately). In contrast to the leapfrog
and Quinn et al.\ integrators, the trajectory of a particle in SEI is
not a straight line between kick steps. This might make collision
detection harder. However, the curved trajectories are a real feature
of the physics in Hill's approximation. Therefore, it does not make
sense to choose an integrator that solves the equations of motion
incorrectly just to search for collisions along those incorrect
trajectories exactly -- it is better to detect collisions
approximately along exact trajectories than the reverse. Developing an
efficient collision algorithm for curved trajectories of this kind is
a research problem that needs further work.  An obvious first step is
to re-use the already implemented collision detection algorithms by
approximating the trajectory as the line that joins the initial and
final positions defined by the curved trajectory. This should work
reasonably well so long as $\Omega \Delta t\ll 1$.

\subsection{Symplectic epicycle-Kepler integrator (SEKI)}\label{sec:wheei}

The integrator described in the previous subsection is designed for
the case where the forces due to the potential $\Phi$ are small
compared to the forces that govern the epicyclic motion. We now
describe an integrator for a situation in which the force due to the
Kepler potential $\Phi(\mathbf{r})=-Gm/|\mathbf{r}|$ is comparable to
or stronger than the forces governing the epicyclic motion.

We first note that one can integrate motion in the Kepler Hamiltonian
$H_{\text{Kep}}(\mathbf{r},\mathbf{p})\equiv
H_{\text{Kin}}(\mathbf{p})+\Phi(\mathbf{r})=\frac12\mathbf{p}^2-Gm/|\mathbf{r}|$
exactly up to machine precision. Efficient methods for doing so are described by
\cite{WisdomHolman1991}. Also note that one can rewrite
Eq.~(\ref{eq:hamiltonian}) as
\begin{align}
H(\mathbf{r},\mathbf{p}) &= H_0(\mathbf{r},\mathbf{p})+H_{\text{Kep}}(\mathbf{r},\mathbf{p}) -H_{\text{Kin}}(\mathbf{p}) .
\end{align}
This motivates the following scheme, which we call symplectic epicycle-Kepler integrator (SEKI):
\begin{align}
& \hat H_{\text{SEKI}}(\Delta t) = \label{eq:wheei} \\
& \quad\quad\hat H_0(\tfrac12{\Delta t})
\;\hat H_{\text{Kin}}(-\tfrac12{\Delta t})
\;\hat H_{\text{Kep}}(\Delta t)
\;\hat H_{\text{Kin}}(-\tfrac12{\Delta t})
\;\hat H_0(\tfrac12{\Delta t}). \nonumber
\end{align}
Note that the drift operator, $\hat H_{\text{Kin}}$, has a negative
time-step. This scheme is symplectic, second-order, and
time-reversible. These statements are also true for other symmetric
permutations of these operators. This particular permutation has been
chosen because the computationally most expensive operator
$H_{\text{Kep}}$ is called only once per time-step. Of course, if
output is not needed at every time-step the half-steps at the end of
step $n$ and the start of step $n+1$ can be combined; then the
alternative scheme
\begin{equation}
\hat H_{\text{Kep}}(\tfrac12{\Delta t})
\;\hat H_{\text{Kin}}(-\tfrac12{\Delta t})
\;\hat H_0(\Delta t)
\;\hat H_{\text{Kin}}(-\tfrac12{\Delta t})
\;\hat H_{\text{Kep}}(\tfrac12{\Delta t})
\end{equation}
is hardly more expensive. 

Note that these operators are defined in $(\mathbf{r},\mathbf{p})$
phase space. For the operators $\hat H_{\text{Kin}}$ and $\hat
H_{\text{Kep}}$ the canonical momentum $\mathbf{p}$ is equal to the
velocity $\mathbf{v}=\dot{\mathbf{r}}=\partial H/\partial\mathbf{p}$ but
for $\hat H_0$ we have $\mathbf{v}=\mathbf{p}+\Omega
\mathbf{r}\cross\mathbf{e}_z$.

\section{Tests}\label{sec:tests}

We ran many test integrations to study the convergence, accuracy, and
computational cost of the integrators presented in the previous
section.  We present only four representative examples.  All of these
tests use the algorithm in Appendix \ref{app:rot} to minimize
round-off errors. Without loss of generality, we set $\Omega=1$ from
now on.

\subsection{Epicyclic motion} \label{sec:epi}
\begin{figure}
\centering
\includegraphics[width=\columnwidth]{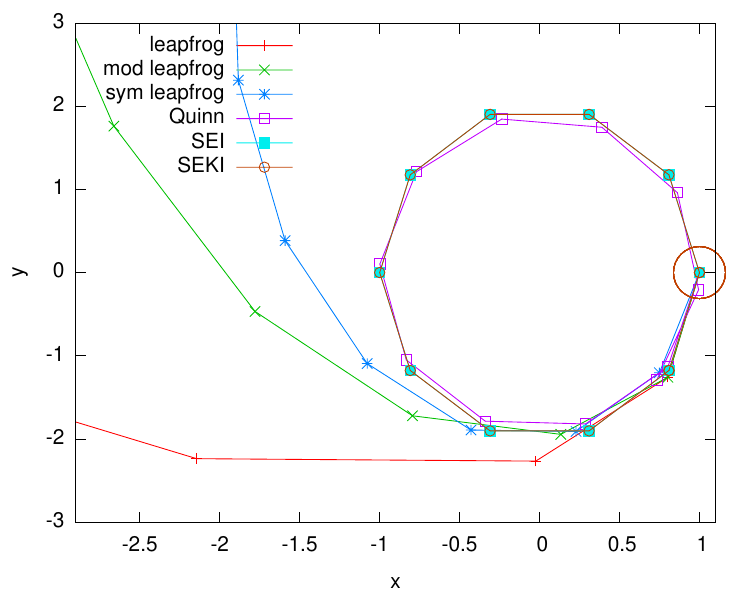}
\caption{Particle motion during one epicycle using different
  integrators and a time-step $\Delta t=0.1\cdot
  2\pi\Omega^{-1}$. The initial position, which is also the
  exact final position,  is marked by a
  circle. \label{fig:orbit_test1}}
\end{figure}
\begin{figure*}
\centering
\captionsetup{margin=2ex}
\subfloat[Relative energy error as a function of time-step. \lisb]{\includegraphics[width=0.5\textwidth]{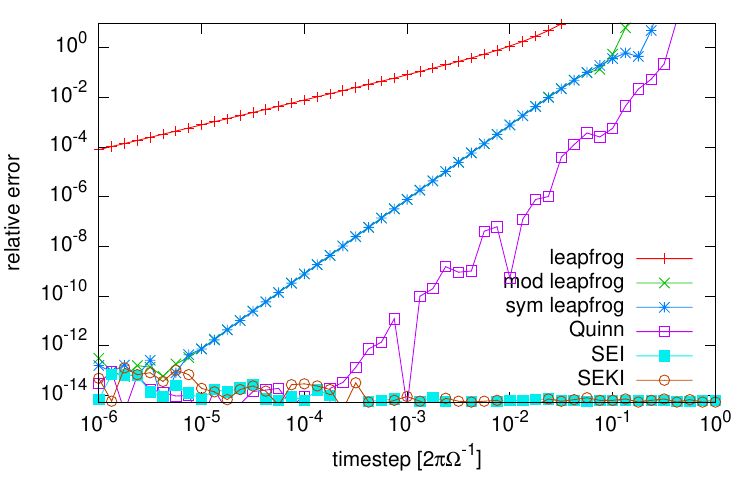}\label{fig:error_test1}}
\subfloat[Computation time as a function of relative energy error. \lisb]{\includegraphics[width=0.5\textwidth]{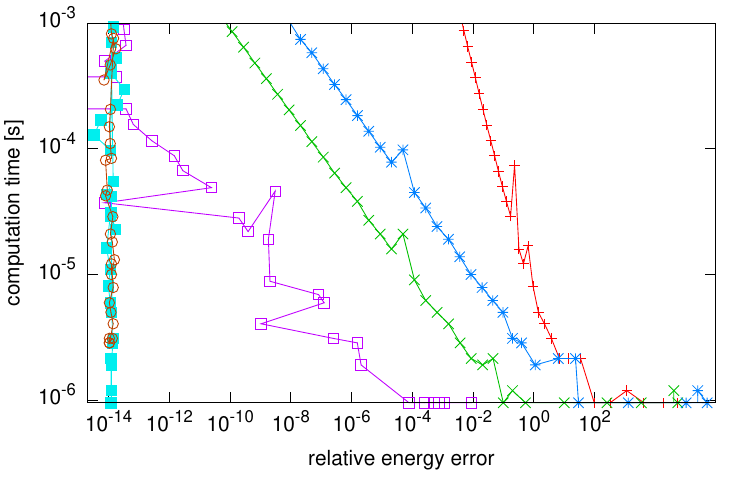}\label{fig:computationtime_test1}}
\caption{Epicyclic motion with no perturbing forces (Sect.\ \ref{sec:epi}). In this case SEI and SEKI follow the
  motion exactly. The method described in the Appendix was used to
  control round-off errors. The integration time is one epicycle
  period.\label{fig:test1}}
\end{figure*}

We first examine the case in which the perturbing force $\mathbf{f}$
in Eq.~(\ref{eq:hill}) vanishes, which corresponds to $m\rightarrow 0$ or
$r_{\text{Hill}}\rightarrow0$. In this case Hill's equations of motion
can be solved analytically and lead to epicyclic motion. We initialize
the particle position to $\mathbf{r}=(1,0,0)^T$ and the velocity to
$\mathbf{\dot r}=(0,-2,0)^T$, which corresponds to a trajectory that
is a clockwise closed ellipse centered on the origin.  We integrate the
trajectory forward in time for one epicycle period
($t=2\pi\Omega^{-1}$).

To illustrate the behavior of each integrator, we use a relatively
large time-step (one tenth of the epicycle period) and plot the position
of the particle at every time-step in Figure~\ref{fig:orbit_test1}.  As
expected,  SEI and SEKI follow the analytic solution exactly. The
Quinn et al.~integrator yields an approximate ellipse but exhibits a
phase error of $6^\circ$ per epicycle period. All three versions
of the leapfrog integrator diverge badly from the 
analytic solution in less than one-quarter of an epicycle period.

In Figure~\ref{fig:error_test1} we plot the relative energy error as a
function of the time-step. As expected, SEI and SEKI are exact to
machine precision for all time-steps.  The Quinn et al.~integrator
and the modified and symmetrized leapfrog integrators converge
quadratically until machine precision is reached. Leapfrog converges
only linearly.  In Figure~\ref{fig:computationtime_test1} we plot the
computation time as a function of the relative error.  This plot shows
the fastest integrator for a desired precision.  Of course, in this
test case the choice is trivial, as SEI and SEKI give the exact
solution for any time-step.

\subsection{Perturbed epicyclic motion} \label{sec:epip}
\begin{figure*}
\centering
\captionsetup{margin=2ex}
\subfloat[Relative energy error and phase error as a function of time-step. \lisb]{\includegraphics[width=0.5\textwidth]{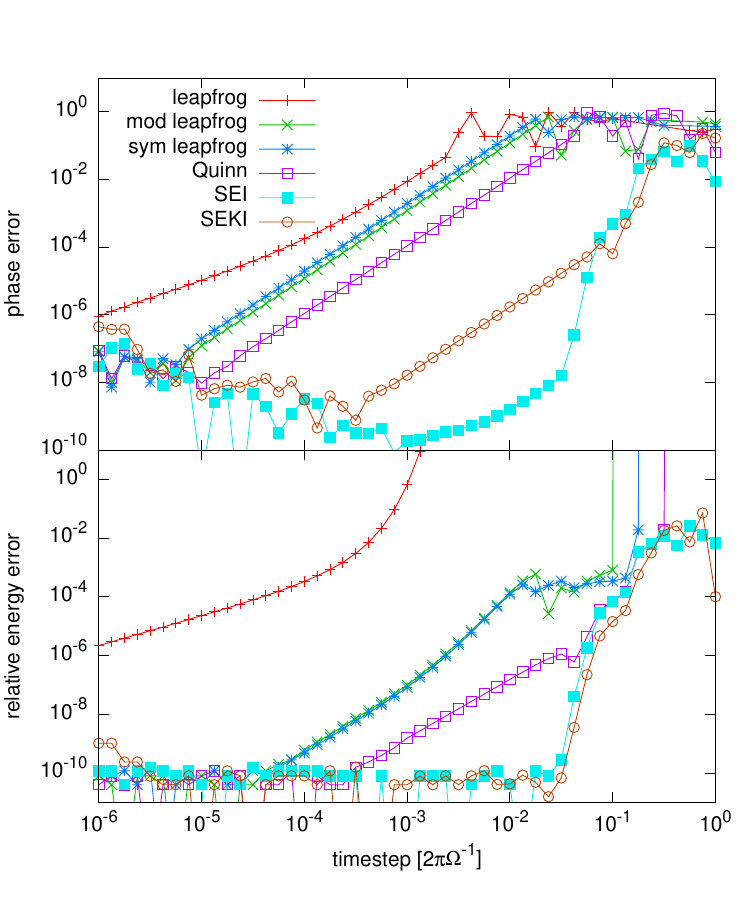}\label{fig:error_test2}}
\subfloat[Computation time as a function of relative energy error and phase error. \lisb]{\includegraphics[width=0.5\textwidth]{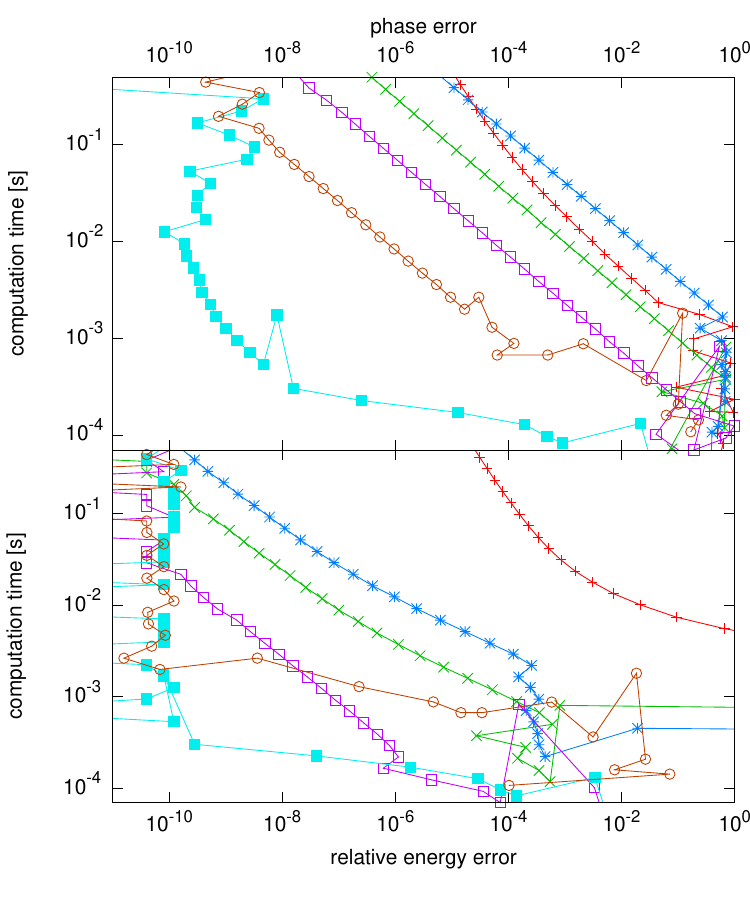}\label{fig:computationtime_test2}}
\caption{Perturbed epicyclic motion (Sect.\ \ref{sec:epip}). The trajectory passes the
  perturbing point mass at an impact parameter of
  $8\,r_{\text{Hill}}$. The integration time is 100 epicycle
  periods.\label{fig:test2}}
\end{figure*}

The motion of a test particle in the presence of a mass $m\neq0$ can
be described by a perturbed epicycle when the mass is sufficiently
small, or, in other words, when the test particle is sufficiently far
away from the mass or moving sufficiently fast, as quantified by Eq.~(\ref{eq:approx}).
We place the particle initially on a circular
orbit around mass $M$ (uniform motion along $\mathbf{e}_y$ in Hill's
approximation), with positions and velocities
$\mathbf{r}=(5.55,2613.91,0)^T$ and $\mathbf{\dot r}=(0,-8.32,0)^T$.
The trajectory passes the perturbing mass at an impact parameter
corresponding to $8\,r_{\text{Hill}}$. The integration time is 100
epicycle periods. As an astrophysical example of such trajectories, we refer the reader
to a study of the stochastic motion of moonlets embedded in 
Saturn's rings by \cite{ReinPapaloizou2010}.

In Figure~\ref{fig:test2} we plot the same diagnostics as in
Figure~\ref{fig:test1} and additionally the phase error. We also
looked at other measures of the accuracy of the integrators but do not
show the results. SEI (for unbound orbits) and SEKI (for bound orbits,
see below) perform at least as well as the other integrators tested by
all measures that we have examined, and much better by most of
them. SEI and SEKI also turn out to be exceptionally good at calculating
the phase of the epicyclic motion. 

One can see from Figure~\ref{fig:test2}a that SEI and SEKI exhibit
energy errors that are up to three orders of magnitude smaller than
any other integrator for typical time-steps used in most simulations
($\Delta t \sim 10^{-3}$--$5\cdot10^{-2})$.  Eventually the errors are
dominated by round-off errors ($\Delta t\lesssim
5\cdot10^{-4}$) for all integrators.  One can further see that at a
fixed time-step SEI produces a phase error that is up to seven~(!)
orders of magnitude smaller than the error produced by the Quinn et
al.\ integrator.  From Figure~\ref{fig:test2}b it is clear that SEI is
also by far the fastest integrator for a given precision. 

\subsection{Strongly perturbed epicyclic motion} \label{sec:sepip}
\begin{figure*}
\centering
\captionsetup{margin=2ex}
\subfloat[Relative energy error and phase error as a function of time-step. \lisb]{\includegraphics[width=0.5\textwidth]{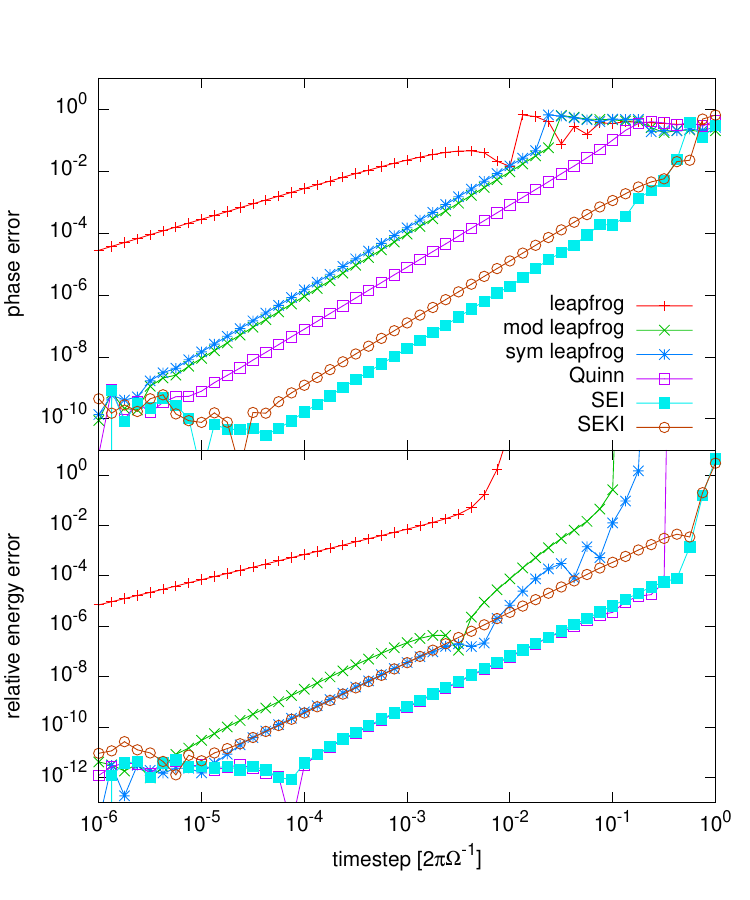}\label{fig:error_test2bcombined}}
\subfloat[Computation time as a function of relative energy error and
phase error. \lisb]{\includegraphics[width=0.5\textwidth]{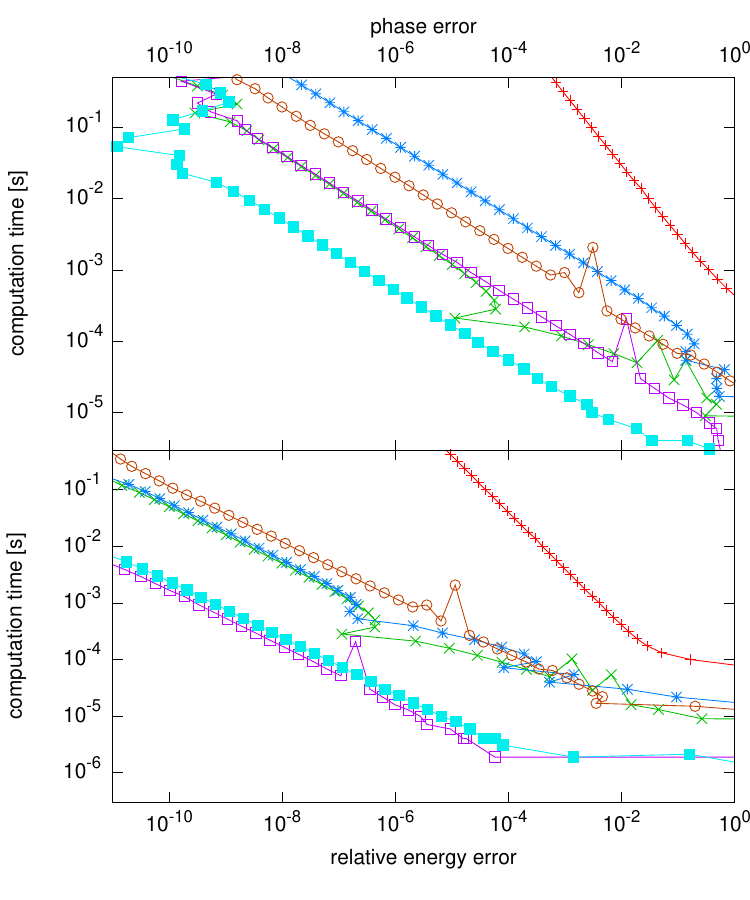}\label{fig:computationtime_test2bcombined}}
\caption{Strongly perturbed epicyclic motion (Sect.\ \ref{sec:sepip}). The trajectory has an impact parameter of
  $1\,r_{\text{Hill}}$ corresponding to a horse-shoe orbit. The integration time is 100 epicycle
  periods.\label{fig:test2b}}
\end{figure*}

We also performed tests in which we place the test particle on an
orbit with an impact parameter of $1r_\text{Hill}$, so the perturbing
forces are much stronger than in the previous case. The trajectory of
the test particle is then a horse-shoe orbit.  

In Figure~\ref{fig:test2b} we plot the same diagnostics as in
Figure~\ref{fig:test2} for this case. One can see that SEI and the
Quinn et al.\ integrator perform equally well as measured by the
energy error. However, the phase error is more than two orders of
magnitude smaller using the SEI or the SEKI integrator rather than for
any of the other integrators.  Note that the SEKI integrator has
significantly larger energy error than SEI. This is because for this test case all
of the Hamiltonian operators in Eqs. (\ref{eq:hamsei}) and
(\ref{eq:wheei}) are roughly of the same magnitude and
thus, the commutators of the operators that give rise to large 
integration errors. Because SEKI is based on a split into
five operators while SEI is based on three, there are more commutators and the total
error is larger in SEKI. 

\subsection{Perturbed Keplerian motion} \label{sec:pkm}

\begin{figure*}
\centering
\captionsetup{margin=2ex}
\subfloat[Relative energy error as a function of time-step. \lisb]{\includegraphics[width=0.5\textwidth]{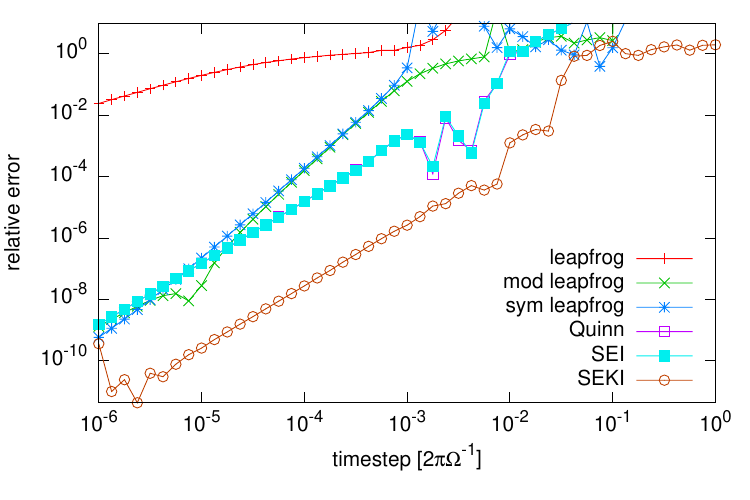}\label{fig:error_test3}}
\subfloat[Computation time as a function of relative error. \lisb]{\includegraphics[width=0.5\textwidth]{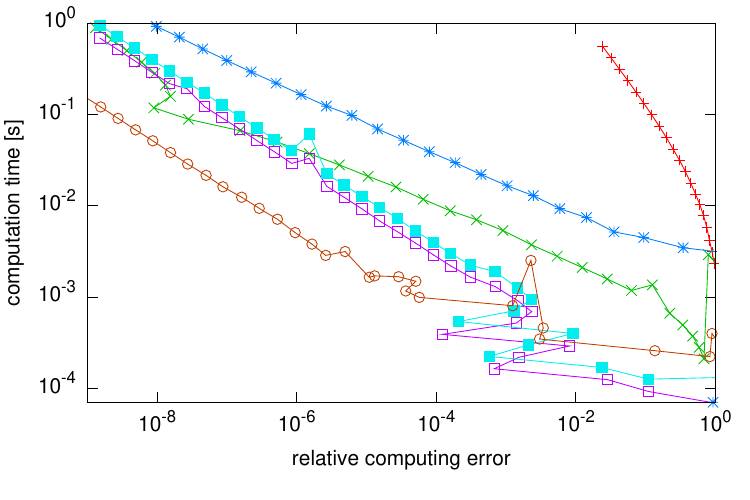}\label{fig:computationtime_test3}}
\caption{Perturbed Keplerian motion (Sect.\ \ref{sec:pkm}). The initial orbit is circular
  with semi-major axis $0.18\,r_{\text{Hill}}$. The integration time
  is 10 epicycle periods or 226 orbital periods around the mass
  $m$. \label{fig:test3}}
\end{figure*}

\begin{figure}
\centering
\includegraphics[width=\columnwidth]{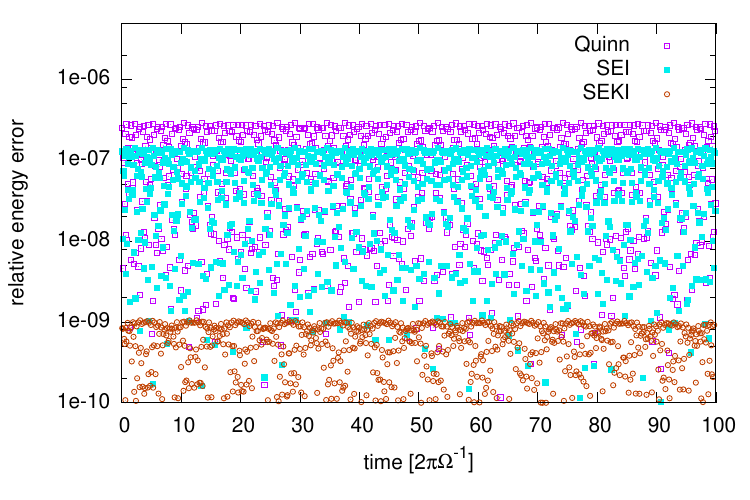}
\caption{Relative energy error using different integrators and a
  time-step $\Delta t=10^{-5}\; 2\pi\Omega^{-1}$. The algorithm
  described in Appendix \ref{app:rot} was used to minimize round-off errors.\label{fig:energy}}
\end{figure}

We finally test the strong gravity regime, where the forces that
govern the epicyclic motion can be viewed as a perturbation to a
Keplerian orbit.  The initial positions and velocities are
$\mathbf{r}=(0.125,0,0)^T$ and $\mathbf{\dot r}=(0,-0.354,0)^T$, which
correspond to an initially circular, bound orbit of $m_1$ and
$m_2$. The semi-major axis of this orbit corresponds to $0.18
r_{\text{Hill}}$.  The integration time is ten epicycle periods,
corresponding to 226 Kepler orbital periods. 

In Figure~\ref{fig:test3} we plot again the same diagnostics as in
Figure~\ref{fig:test1}. 
The SEKI integrator is the most robust integrator in
the sample, with the energy error at a given time-step more than two orders of
magnitude smaller than its competitors. This performance comes with a
drawback, as each time-step is computationally more expensive;
nevertheless, at fixed energy error it is still more than one order of
magnitude faster than its competitors. Its relative advantage 
improves further for orbits with smaller semi-major axis (relative to~$r_{\text{Hill}}$).

Also note that for integrations in which the evaluation of forces is
expensive (for example in a tree code), the SEKI integrator has a further
advantage as it uses fewer time-steps.

Figure~\ref{fig:energy} shows the time evolution of the relative energy
error in the integration with $\Delta t=10^{-5}\; 2\pi\Omega^{-1}$.
All integrators (except the various flavors of leapfrog which are far off the scale and
therefore not plotted) show no sign of a linear drift in the energy
error: the maximum error is independent of time. This
good behavior is due to the symplectic nature of these integrators as
well as the use of the procedure described in Appendix~\ref{app:rot}
to control round-off error.  The SEKI integrator is more than two
orders of magnitude better than any other integrator in this example.

\section{Conclusions}\label{sec:conclusions}

We have presented two new integrators for studying the small-scale
dynamics of disks using Hill's approximation.

The first is a simple symplectic and time-reversible integrator for
Hill's equations of motion that we call \textit{symplectic epicycle
  integrator (SEI)}.  In the absence of small-scale forces such as the
self-gravitational forces of the disk particles, SEI solves Hill's
equations of motion (epicyclic motion) exactly; otherwise the error
over a fixed time integral scales as $O(\epsilon\,\Delta t^2)$ when
the small-scale forces are $O(\epsilon)$.  Numerical tests using a
variety of measures have shown that SEI always converges much faster
than various flavors of leapfrog, and often much faster than the \cite{Quinn2010}
integrator. For small~$\epsilon$ ($\epsilon\sim 3(r_{\text{Hill}}/r)^3\sim 0.01$) the phase error can be
up to seven orders of magnitude smaller than the phase error from the Quinn et
al.\ integrator at the same time-step (Figure~\ref{fig:test2}a). The
computational cost per time-step of SEI and the Quinn et al. integrator are equivalent.
Although SEI is simple to code, a C~implementation can be
downloaded at \url{http://sns.ias.edu/~rein/}.

The second integrator is also symplectic and time-reversible and is
called the \textit{symplectic epicycle-Kepler
  integrator (SEKI)}. This integrator is useful in following bound
two-body orbits in the sheared sheet and in this case can yield errors
that are several orders of magnitude smaller than SEI. 

These integrators can be generalized in several ways. First, higher
order integrators can be constructed by concatenating SEI and SEKI
steps of varying lengths \citep[e.g.,][]{Yoshida1993}. Second,
although the discussion in this paper has, for simplicity, focused on
the three-body problem, SEI can be applied to the
$N$-body problem using shear periodic boundary conditions
\citep[e.g.,][]{Richardson1994}. The force $\mathbf{f}$
in Eq.~(\ref{eq:hill}) can also describe gas drag on particles,
although in this case the dynamics is not described by a Hamiltonian
so the advantage of a symplectic, time-reversible integrator is less
clear. 

The SEKI integrator can also be generalized beyond Hill's
approximation. For example, consider a test particle
orbiting in the gravitational field of a binary star with masses $M_1$ and $M_2$.  The
Hamiltonian can be written as a sum of two Keplerian Hamiltonians
minus the kinetic energy, 
\begin{align}
H\left(\mathbf{r},\mathbf{p}\right) = H_{\text{Kep,}M1}\left(\mathbf{r},\mathbf{p}\right) +  H_{\text{Kep,}M2}\left(\mathbf{r},\mathbf{p}\right) - H_{\text{Kin}}\left(\mathbf{p}\right). 
\end{align}
The first two terms can be solved exactly, and the
last term is simply a drift. Thus, the analog of Eq.~(\ref{eq:wheei})
provides a second-order accurate, symplectic and time-reversible
integrator that is exact when the test-particle motion is dominated by
the gravitational field from either body. 


\section*{Acknowledgments}
  This work was supported in part by NSF grant AST-0807444 and by NASA
  grant NNX08AH83G. We thank Tobias Heinemann for helpful discussions
  and Tom Quinn for a thoughtful and constructive referee's report.


\bibliographystyle{mn2e}
\bibliography{full}

\appendix
\section[]{Round-off error due to rotations}\label{app:rot}

Floating-point operations on a computer are subject to rounding
errors.  The IEEE standard for floating-point arithmetic~(IEEE~754)
specifies the rounding algorithm and ensures that the rounding error
for additions and multiplications is quasi-random.  Thus, the rounding
error should grow with time as $O(\machineepsilon\cdot(t/\Delta t)^{1/2})$,
where $\machineepsilon$ is the machine precision.

Both the SEI and SEKI integrators use rotations, which are written in
Eqs.~(\ref{eq:SEI2}) and (\ref{eq:SEI3}) as a rotation matrix with
angle $\phi =\Omega\Delta t$. However, operating with this matrix on a position
vector $(x,y)$ will not in general preserve the norm
$r=(x^2+y^2)^{1/2}$ because numerically $\sin^2 \phi + \cos^2\phi \neq
1$.  Since $\phi$ is the same at every time-step, the error grows
linearly with the number of operations, $O(\machineepsilon\cdot t)$, much worse
than the $O(\machineepsilon\cdot t^{1/2})$ behavior described above \citep[][and references
therein]{Petit1998}. This results in an undesirable linear drift in energy.

For other integrators that solve the Kepler problem using rotations,
such as the Wisdom--Holman integrator, this problem is usually not
important. This is because the rotation angle in the Kepler problem
depends implicitly on radius (Kepler's law) and thus the rotation is
actually a so-called twist map. There exists a KAM-like theorem
\citep{Blank1997} for twist maps that restricts the solution to an
invariant torus in phase space.

\cite{Petit1998} describes one way to solve this problem by
decomposing the rotation operator in Eq.~(\ref{eq:SEI2}) into three
shear operators
\begin{align}
&\left( \begin{array}{cc}
\cos\phi & \sin\phi \\
-\sin\phi & \cos\phi \end{array} \right)=\nonumber\\
&\quad\quad \left( \begin{array}{cc}
1 & 0 \\
-\tan\tfrac12 \phi & 1 \end{array} \right)\cdot 
\left( \begin{array}{cc}
1 & \sin\phi \\
0 & 1 \end{array} \right)\cdot
\left( \begin{array}{cc}
1 & 0 \\
-\tan\tfrac12 \phi  & 1 \end{array} \right).\label{eq:twist}
\end{align}
Why does this help? Suppose we replace the rotation matrix in Eq.~(\ref{eq:SEI2}) 
by an arbitrary matrix $\bf R$. It is then
straightforward to show that the transformation from
$(x^n,y^n,p_x^n,p_y^n)^T$ to $(x^{n+1},y^{n+1},p_x^{n+1},p_y^{n+1})^T$
defined by Eqs.~(\ref{eq:epia}) to (\ref{eq:fff}) is symplectic if
and only if $\mbox{det}\,{\bf R}=1$. Since 1 and 0 are represented
exactly in floating-point arithmetic, each of the matrices on the
right side of Eq.~(\ref{eq:twist}) has a determinant of exactly 1. Thus
the transformation is symplectic whether or not $\sin\phi$ and
$\tan\half\phi$ are related by the appropriate trigonometric
identity, and hence is insensitive to round-off errors in evaluating
these functions.  

We tested both implementations of the rotation operator,
Eq.~(\ref{eq:SEI2}) and Eq.~(\ref{eq:twist}). As expected, the
implementation using shear operators shows no sign of a linear drift
in energy (Figure~\ref{fig:energy}). In contrast, for the test case
presented in Figure~\ref{fig:energy}, the straightforward implementation
of Eq.~(\ref{eq:SEI2}) produces a linear drift of about $10^{-7}$
after only 100 epicycle periods.  The additional computational cost of
implementing the rotations using shear operators is negligible, and in
long integrations this refinement can dramatically improve the
accuracy.

\label{lastpage}

\end{document}